# Optical Control of Fluorescence Spatial Hole Burning Effect in Monolayer $WS_2$

Yichun Pan(潘议淳)[1], Liqing Zhu(朱莉晴)[1], Zheng Wang(王政)[1*], Weihang Zhou(周伟航)[1*]

[1]Wuhan National High Magnetic Field Center and School of Physics, Huazhong University of Science and Technology, Wuhan 430074, China

Supported by the National Natural Science Foundation of China under Grant No. 12274159 and National Key Research and Development Program of China under Grant No. 2022YFA1602700.
*Corresponding author. Email: wangzheng@hust.edu.cn; zhouweihang@hust.edu.cn

Doping plays a crucial role in both electrical and optical properties of semiconductors. In this work, we report observation, as well as optical control, of fluorescence spatial hole burning effect in monolayer $WS_2$. We demonstrate that the pronounced exciton-exciton annihilation process, in combination with the efficient capture of holes by intrinsic sulfur vacancy defects, induces significant photo-doping effect and eventually leads to fluorescence spatial hole burning. By means of a dual-beam pumping fluorescence imaging technique, we reveal that the recovery process of the spatial hole burning effect exhibits a double-exponential behavior. The fast recovery process originates from the release of trapped holes under the illumination of the probe beam, while the slow process corresponds to the re-adsorption of electronegative gas molecules. Surprisingly, our results demonstrate that the electronegative sulfur vacancies can achieve ultralong-term storage of holes. Moreover, both the release and the rate of release of holes can be fully controlled by laser irradiation. These findings demonstrate the great potential of transition metal dichalcogenides in the development of optically-control excitonic devices.

## Introduction

It is nowadays well-known that doping is indispensable for optimizing and tailoring the electrical and optical properties of semiconductors, enabling the development of a wide range of modern electronic and optoelectronic devices.[1,2] As typical two-dimensional van der Waals semiconductors, doping of transition metal dichalcogenides (TMDCs) has received tremendous attention in recent years. [3–6] Several techniques have been developed for the efficient doping of TMDCs, including electrostatic doping, chemical doping, physical doping and photoinduced doping.[7–9] In particular, photo-doping offers non-destructive, reversible, and even real-time modulation of both electrical and optical properties of TMDCs. Thanks to the efforts devoted in the past decade, several key issues concerning photo-doping of TMDCs, such as the role of adsorbed electronegative gas molecules, the effect of photo-oxidation and laser-induced chalcogen vacancies, have been clarified.[10–13] Undoubtedly, these advancements have provided people with a comprehensive understanding of the formation mechanism of photo-doping. However, it must be pointed out that these reports mainly focus on the physical and chemical doping processes caused by laser irradiation, while very little attention has been paid to the recovery process of photo-doping and its underlying physics.

On the other hand, as typical two-dimensional materials, one of the most distinctive properties of monolayer TMDCs is their pronounced excitonic effect. Due to the tight quantum confinement and the reduced Coulomb screening effect, excitonic binding energies of monolayer TMDCs are reported to be up to several hundreds of meV.[6,14–18] As a result, optical properties of TMDCs are essentially governed by excitons under both low-temperature and room-temperature conditions. Moreover, due to the existence of two degenerate valleys K and K′, monolayer TMDCs are widely known to possess a group of complicated excitonic states.[19–28] Spectroscopic studies have shown that emission from TMDCs usually has contributions from neutral excitons, charged excitons, biexcitons, bound excitons, as well as other relevant species.[29–31] Considering their two-dimensional geometry, a thorough understanding of the interactions among these excitonic species would be critical for the understanding of the photo-doping mechanism of monolayer TMDCs. However, the understanding of the role of excitonic interactions in photo-doping, especially in the recovery process, remains quite limited so far.[32]

In this work, we report systematic studies on photo-doping and its optical control in monolayer $WS_2$ flakes. We found that the intrinsic sulfur vacancy defects can efficiently capture holes generated by the exciton-exciton annihilation process. This induces a significant photo-doping effect and eventually leads to fluorescence spatial hole burning. By means of a time-resolved dual-beam pumping fluorescence imaging technique, we were able to capture the whole picture of the fluorescence spatial hole burning effect, including its generation and recovery processes. In particular, we reveal that the recovery process consists of two distinct processes with significantly different time constants. The fast process was confirmed to originate from the release of trapped holes under the illumination of a weak probe beam, while the slow process stems from the re-adsorption of electronegative gas molecules. Based on our time-, space-, and energy-resolved spectroscopic analyses, ultralong-term storage of holes and its full optical control were also revealed.

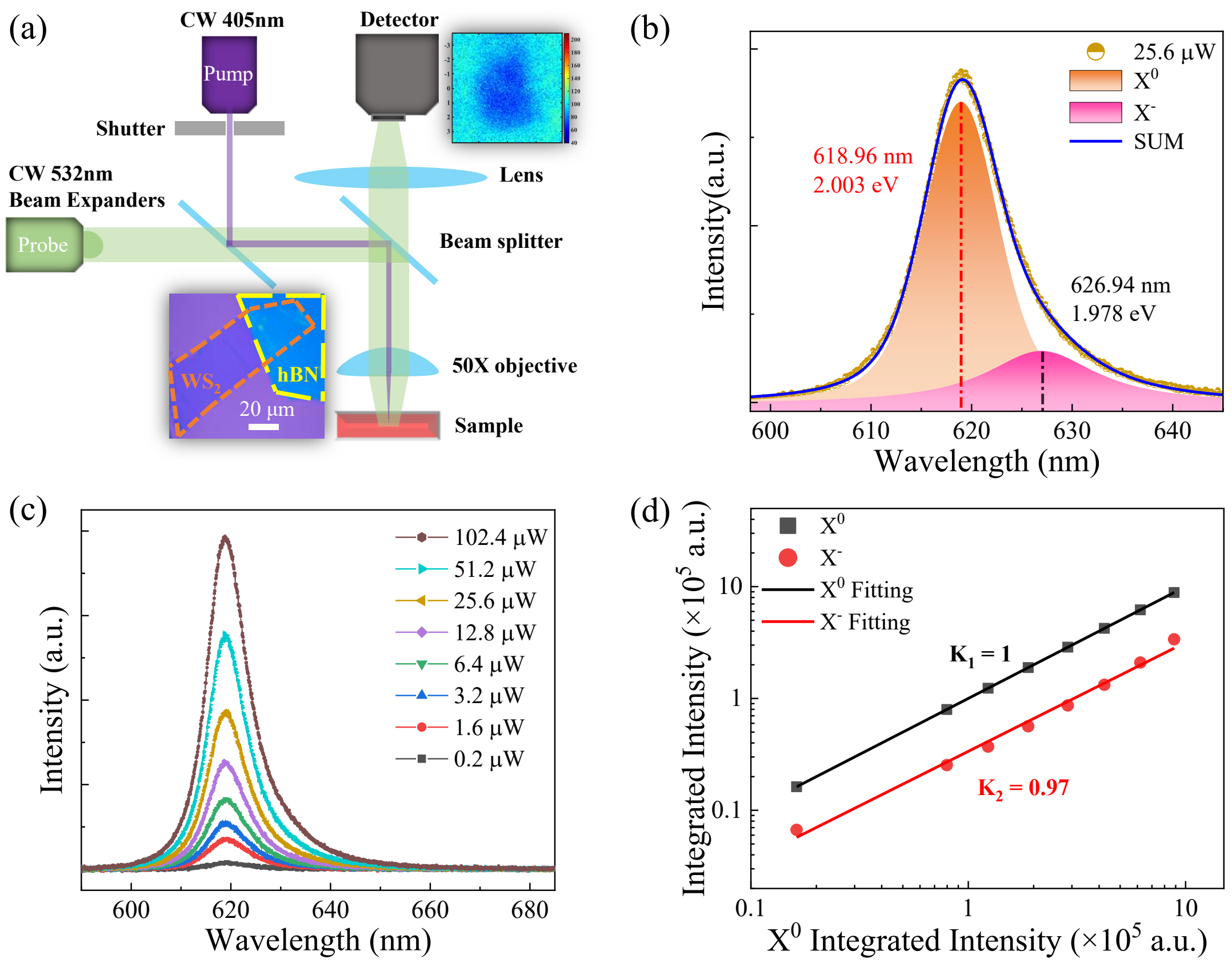

**Fig. 1. Experimental setup and spectral analyses for monolayer $WS_2$.** (a) Schematic diagram of the dual-beam pumping fluorescence imaging system; (b) Lorentz fitting of typical PL spectrum of $WS_2$. (c) Power-dependent PL spectra of $WS_2$; (d) Intensities of the fitted $X^0$ and $X^-$ peaks as a function of the integrated intensity of $X^0$. The solid lines are fitting curves using the function $y = a + x^k$.

## Results

The samples we used are monolayer $WS_2$ flakes mechanically exfoliated from $WS_2$ single crystals (HQ, Netherlands) and transferred onto clean $SiO_2$ substrates. A portion of the flakes were covered by a thin layer of hexagonal boron nitride (hBN) to serve as the control group in our experiments, as shown by the inset in Fig. 1(a). A schematic diagram of our home-built dual-beam pumping fluorescence imaging system is also presented in Fig. 1(a) for clarity. A relatively high-power 405 nm continuous-wave (CW) laser, focused onto sample surface using an Olympus 50X objective lens and with a spot diameter of ~3 μm, was used as the pumping beam. A relatively low-power 532 nm CW laser, focused onto samples via the same objective lens, was used as the probe beam. The probe beam was expanded using a beam expander and has a large spot diameter of ~500 μm on the sample surface.

Unless otherwise specified, power of the pumping laser beam in our experiments was set to be ~3 mW, corresponding to an estimated exciton density of ~$2.6\times10^{11}$ cm$^{-2}$ (assuming an absorption coefficient of 10% and an exciton lifetime of approximately 30 ps ). This value is far below the threshold exciton concentration for Mott transition, which is estimated to be ~$10^{13}$ cm$^{-2}$. According to literatures, for exciton density ranging from $10^{9}$ ~ $10^{12}$ cm$^{-2}$ in monolayer $WS_2$ or $MoS_2$, significant exciton-exciton annihilation (EEA) effect, which is the excitonic version of Auger recombination, can be readily observed.[33–35] In our discussion below, we will also show that EEA plays a crucial role in the photo-doping and fluorescence spatial hole burning effect studied in this work. For each spectral / imaging measurement, exposure time of the pumping laser beam was set to be ~50 ms using a mechanical shutter. To avoid unwanted heating effect, power of the probe beam was typically one to two orders of magnitude lower than the pumping beam. A charge-coupled device (CCD) and a complementary metal-oxide-semiconductor (CMOS) scientific-grade detector were used for time-resolved fluorescence imaging and spectral measurements, respectively, with a sampling rate of 20 Hz.

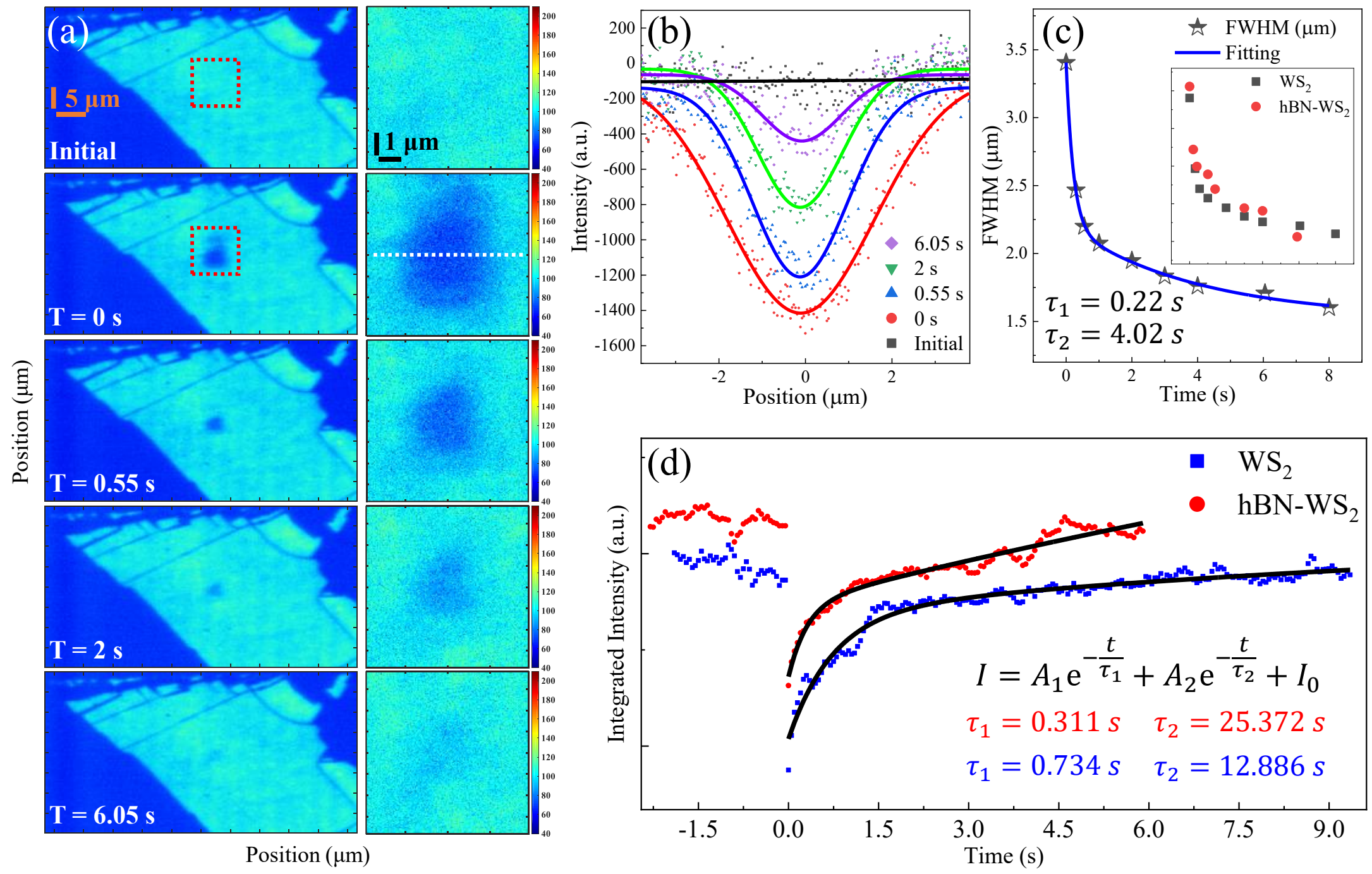

**Fig. 2. Fluorescence spatial hole burning effect and its recovery in monolayer $WS_2$.** (a) Time-resolved fluorescence images of monolayer $WS_2$ flakes measured under our dual-beam pumping scheme. Power of the 405 nm pumping laser: ~8.2 mW. The pumping beam was exposed to the sample for ~50 ms, while the weak probe beam (532 nm) kept exciting the sample during the whole measurement cycle. The right panel shows the zoomed-in images for the region highlighted by the red square in the left. (b) Evolution of the spatial hole burning profile taken along the dotted line in (a), showing the recovery of the spatial hole burning effect. The solid curves are corresponding Gaussian fitting curves. (c) The full-width at half-maximum of the fluorescence hole as a function of time. The blue solid curve is the bi-exponential fitting curve, with the correspond time constants fitted to be $\tau_1$=0.22 s and $\tau_2$=4.02 s, respectively. The inset shows the comparison of the recovery process between bare $WS_2$ and the hBN encapsulated $WS_2$. (d) Integrated PL intensity from the red dashed region in (a) as a function of time. The black solid curves represent bi-exponential fittings. Time constants of the fitting for bare $WS_2$ are $\tau_1$=0.734 s and $\tau_2$=12.886 s, respectively. For hBN-encapsulated flakes, the fitting time constants are $\tau_1$=0.311 s and $\tau_2$=25.372 s, respectively. In all these figures, the zero point in the time domain indicates the moment when the pumping beam was stopped.

Typical photoluminescence spectrum of exfoliated $WS_2$ flakes excited by a 532 nm CW laser is shown in Fig. 1(b). It consists of two peaks, where the neutral exciton ($X^0$) peak locates at ~618.96 nm and the charged exciton peak appears at ~626.94 nm. Fig. 1(c) shows the power dependent PL spectra of $WS_2$. While the pumping power was increased by up to 512 times (from ~0.2 μW to

~102.4 μW), the peak position and the peak profile remain unchanged, suggesting that heating effect for laser power below 102 μW is negligible. To further confirm the assignment of the neutral exciton and charged exciton peaks, we extracted the integrated PL intensity of the $X^-$ peak and plotted it out as a function of the $X^0$ peak intensity, as shown in Fig. 1(d). As one can see, intensity of the $X^-$ peak shows a sub-linear dependence on that of the $X^0$ peak. Based on this feature and the energetic spacing between $X^0$ and $X^-$ in the spectra, we can confirm that $X^0$ and $X^-$ peaks originate from the neutral and charged excitons, respectively.

Fig. 2(a) shows the typical time-resolved fluorescence images of monolayer $WS_2$ flakes measured under our dual-beam pumping scheme. Here, the 405 nm pumping laser beam has a relatively high power of ~8.2 mW. We used it to pump the sample for ~50 ms and then shut it down. On the other hand, the weak 532 nm probe beam kept exciting the sample so that we can measure time-resolved fluorescence images of the sample. The top panel in Fig. 2(a) shows the fluorescence image before the pumping beam was applied. As one can see, the flake shows roughly homogeneous emission, suggesting high quality of the sample. The image denoted by T = 0 s shows the fluorescence image when the sample was excited by the pumping beam for ~50 ms. As demonstrated clearly, emission intensity from the area exposed to the pumping beam (highlighted by the red dotted square) decreases significantly, leading to the formation of fluorescence spatial hole burning. However, about 0.55 s after the pumping beam was shut down, the fluorescence hole narrows obviously, as one can see from the image denoted by T = 0.55 s. This trend continued until the fluorescence hole disappeared completely ~6 s after the pumping beam was stopped. Such evolution can be demonstrated even more clearly in the corresponding zoomed-in images shown in the right panel of Fig. 2(a). To extract more information from the fluorescence spatial hole burning effect, we plotted out the profile of the fluorescence hole at different moments, as shown in Fig. 2(b). The corresponding Gaussian fittings (solid curves) are also given. Obviously, both the width and the depth of the hole decrease gradually. Here, it's interesting to note that the fluorescence hole always has a Gaussian profile, suggesting the probable relationship with the Gaussian-shaped pumping beam. Fig. 2(c) shows the full-width at half-maximum (FWHM) of the fluorescence hole as a function of time, together with its corresponding bi-exponential fitting curve. As demonstrated, the fitting agrees with the experimental data very well, suggesting that the recovery process of the fluorescence hole consists of two channels whose time constants are $\tau_1$ = 0.22 s and $\tau_2$ = 4.02 s, respectively. The inset in Fig. 2(c) shows the comparison of the FWHM evolution between bare $WS_2$ and hBN-encapsulated $WS_2$. In general, both types of samples show bi-exponential behaviors. However, through more careful examination, one could find that time constants for these two types of samples are slightly different. To further confirm the bi-exponential behaviors in the recovery process, we extracted the evolution of the integrated intensity from the region of interest (highlighted by the dotted square in Fig. 2(a)), as shown in Fig. 2(d). Again, evolution of the integrated intensity shows bi-exponential behaviors for both bare $WS_2$ and hBN-encapsulated $WS_2$. Time constants from the fittings are $\tau_1$ = 0.734 s, $\tau_2$ = 12.886 s for $WS_2$ and $\tau_1$ = 0.311 s, $\tau_2$ = 25.372 s for hBN-encapsulated $WS_2$, respectively. The fast process for hBN-encapsulated $WS_2$ is ~2 times faster than that for the bare sample, while its slow process is ~50% slower than that for the bare flake. As an ideal insulator with a wide bandgap, atomically flat surface, high thermal conductivity, but no dangling bonds, the introduction of hBN could not only optimize the defects and sulfur vacancies on $WS_2$ surface but could also effectively prevent the interaction of $WS_2$ with its surrounding atmosphere.[36] As a result, the fluorescence recovery process would be changed. We will discuss this issue further in the following text.

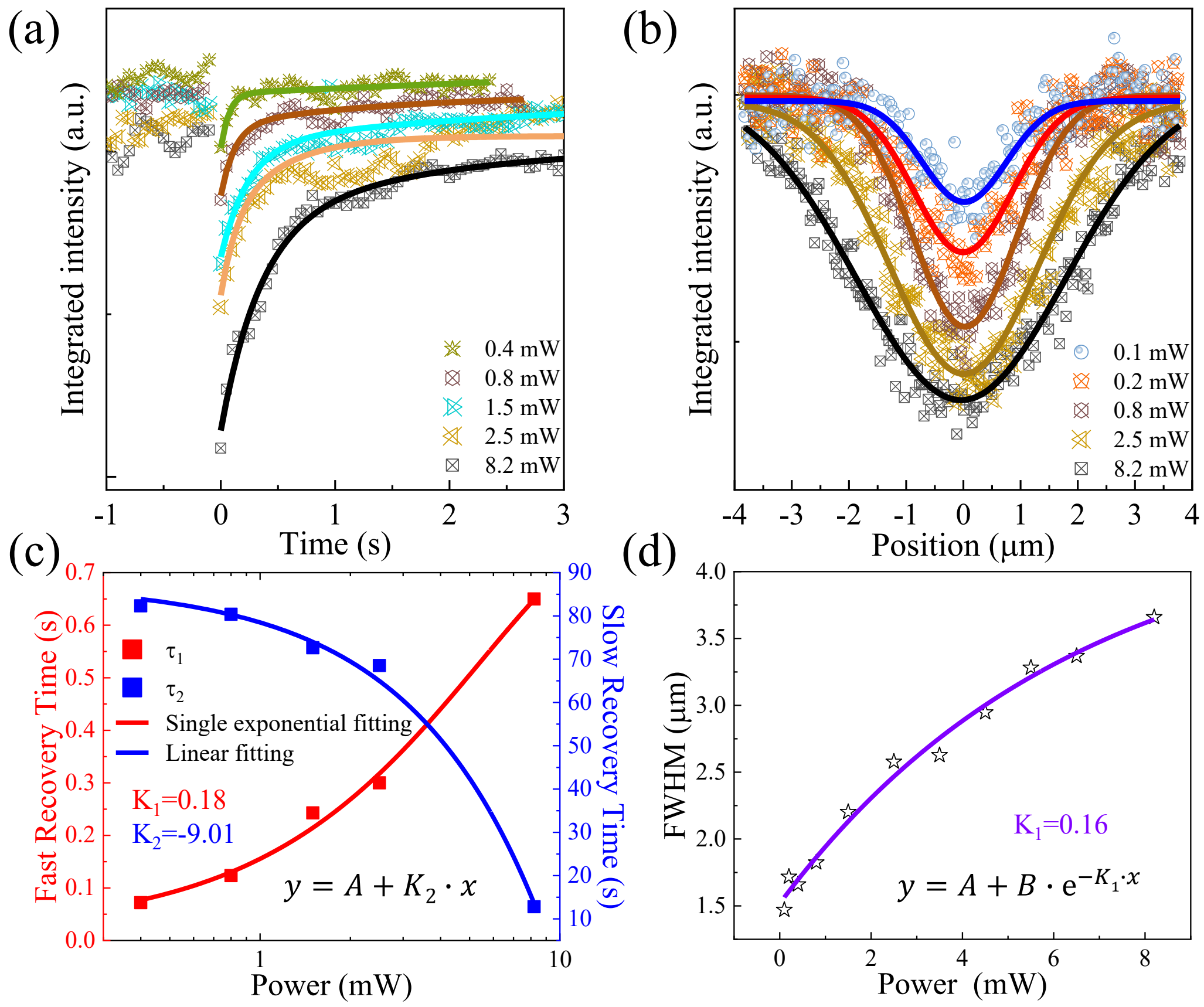

**Fig. 3. Pumping power dependence for the recovery process of the fluorescence spatial hole burning effect.** (a) Evolution of the PL integrated intensity in the time domain for different pumping beam powers. Power of the probe beam is kept unchanged. (b) Profile of the fluorescence spatial hole burning under different pumping beam powers. Power of the probe beam is kept unchanged. (c) Red data points: time constant of the fast recovery process as a function of the pumping beam power. Red solid curve: Yield-fertilizer single-exponential model fitting. Blue data points: time constant of the slow recovery process as a function of the pumping beam power. Blue curve: linear fitting curve. Note that the horizontal axis here uses a logarithmic scale. (d) FWHM of the fluorescence hole as a function of the pumping beam power. The solid curve shows the Yield-fertilizer single-exponential model fitting. Note that the horizontal axis here has been changed back to the linear scale.

To better understand the bi-exponential recovery process of the fluorescence spatial hole burning, we carried out pumping power dependence studies. Fig. 3(a) shows the evolution of the integrated PL intensity under different pumping beam powers (while power of the probe beam remains unchanged). As one can see, quenching of the emission becomes more pronounced as power of the pumping beam was increased. However, the emission intensity always eventually returns to its initial state in a few seconds after the pumping beam was stopped. This tells us that the sample has not been damaged by the high-power pumping beam and that no irreversible chemical reactions have occurred. Fig. 3(b) shows the profile of the fluorescence hole under different pumping beam powers. It's obvious that FWHM of the hole increases monotonically with the pumping power. At the highest pumping power of ~8.2 mW, FWHM of the fluorescence hole (~2 μm) is almost twice the FWHM of the pumping laser spot (~1 μm). This suggests strongly that exciton diffusion is involved in the fluorescence spatial hole burning effect. Fig. 3(c) shows the time constants for the fast (and slow) recovery process, extracted through bi-exponential fitting, as a function of the pumping beam power (while power of the probe beam was kept unchanged). As one can see, time constant for the fast recovery process grows with the pumping power. Moreover, it can be accurately described by the Yield-fertilizer single-exponential model, indicating that it will saturate at high pumping power. In sharp contrast, time constant of the slow recovery process decreases linearly with the pumping power, as highlighted by the blue solid curve fitted using a linear function. These completely different power-dependent behaviors suggests that mechanisms for the fast and slow recovery processes are completely different. Fig. 3(d) shows the FWHM of the fluorescence hole as a function of the pumping beam power. Interestingly, it can be fitted very well by the Yield-fertilizer single-exponential model and saturates at high pumping power, similar to the time constant of the fast recovery process.

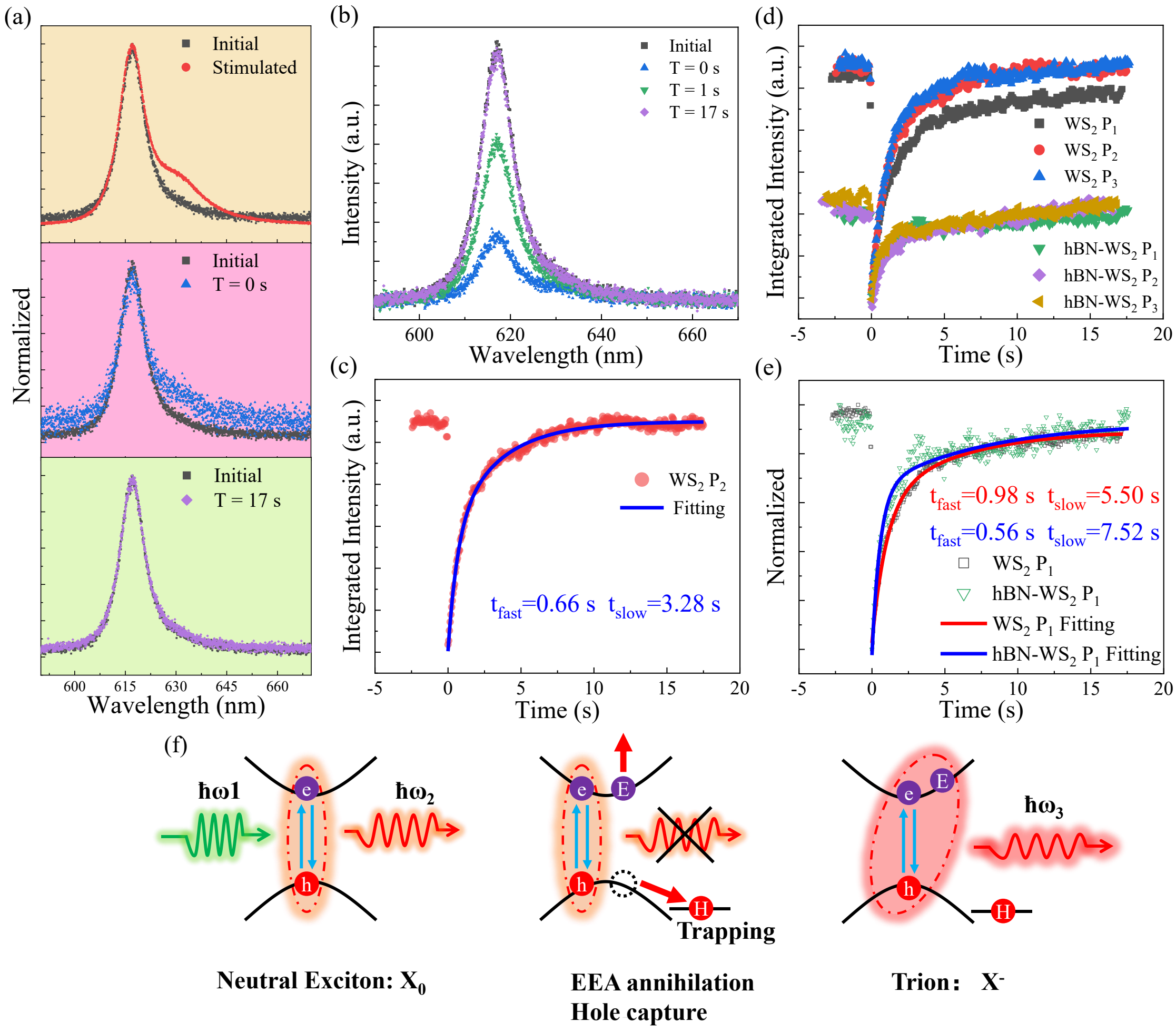

**Fig. 4. Spectral analyses of the fluorescence spatial hole burning and its recovery process in monolayer $WS_2$.** (a) Normalized PL spectra at different moments under our dual-beam pumping scheme. Black solid square (denoted by "Initial"): PL spectra when samples were excited by the probe beam only. Red solid circle (denoted by "Stimulated"): PL spectra when samples were excited by the pump beam and probe beam simultaneously. Blue triangle (denoted by "T = 0 s"): PL spectra taken immediately after the pumping beam was stopped. Pink diamond (denoted by "T = 17 s"): PL spectra taken ~17 s after the pumping beam was stopped. Power of probe beam: ~20 μW; Power of pumping beam: ~3 mW. (b) Typical PL spectra taken at different moments under our dual-beam pumping scheme. (c) Integrated PL intensity of bare $WS_2$ as a function of time. Spot size of the probe beam was changed to be the same as that of the pump beam. Solid curve: bi-exponential fitting curve. Time constants for the fast and slow processes are $t_{fast}$=0.66 s and $t_{slow}$=3.28 s, respectively. (d) Integrated PL intensity as a function of time measured at three different sample positions, for both bare $WS_2$ ($WS_2$ $P_1$, $WS_2$ $P_2$, $WS_2$ $P_3$) and encapsulated $WS_2$ (hBN-$WS_2$ $P_1$, hBN-$WS_2$ $P_2$, hBN-$WS_2$ $P_3$). (e) Comparison of the intensity evolution curve between bare $WS_2$ and hBN-encapsulated $WS_2$. Together shown are the bi-exponential fitting curves (solid curves). Time constants extracted from the bi-exponential fitting are also listed in the figure. (f) Schematic diagram showing the mechanism of the fluorescence spatial hole burning effect.

Till now, we have studied the fluorescence spatial hole burning effect based on the time- and pumping power-dependent behaviors of its integrated PL intensity and FWHM. To further reveal the underlying physics for the fluorescence spatial hole burning effect, especially for its bi-exponential recovery process, it is necessary to analyze the spectral evolution of the samples. As the spatial hole burning effect happens only in the area covered by the pumping beam, only spectra recorded from the pumping beam spot area are meaningful. To improve the spectral signal-to-noise ratio, spot size of the probe beam was reduced to match that of the pumping beam in our spectral measurements. Typical normalized PL spectra of $WS_2$ measured using our dual-beam pumping scheme are plotted in Fig. 4(a). In the top panel of Fig. 4(a), the black data points show the PL spectrum when the sample was excited by the probe beam only, while the red data points show the spectrum when the sample was excited by both the probe and pumping beams simultaneously. As one can see, a new peak in the long-wavelength side emerges when the sample was excited by both probe and pumping beams. Based on the peak position, we can know that this new peak originates from charged excitons. This also suggests that doping concentration was increased when the sample was illuminated by high-power laser beam. In the middle panel of Fig. 4(a), we show the PL spectrum measured immediately when the pumping beam was stopped (the sample was excited by the weak probe beam only). In this case, the charged exciton peak can still be clearly observed. However, we found that intensity of the charged exciton peak decreases gradually after the pumping beam was stopped. In ~17 s after the pumping beam was stopped, the charged exciton peak disappeared and the PL spectrum fully recovers to its initial state, as demonstrated in the bottom panel of Fig. 4(a). Such recovery process can also be identified clearly from the unnormalized spectra, as demonstrated in Fig. 4(b). These phenomena tell explicitly that pumping laser-induced doping will decrease gradually under the illumination by a weak probe beam.

As a comparison, the time-dependent integrated PL intensity, in the case where the probe beam has the same spot size as the pumping beam, is shown in Fig. 4(c). Bi-exponential recovery behaviors can again be clearly observed, thus confirming validity of the experimental results shown in Figs. 2 & 3. To study the effect of hBN encapsulation, typical evolution curves of PL intensity for both bare $WS_2$ and hBN-encapsulated $WS_2$ are plotted together in Fig. 4(d). While both types of samples show bi-exponential recovery behaviors, PL intensity from encapsulated flakes is significantly lower than that from bare $WS_2$. This is because hBN encapsulation isolates $WS_2$ from the atmospheric environment. Such isolation prevents electronegative gas molecules ($O_2$ and $H_2O$) from adsorbing onto $WS_2$ surface, thereby enhancing the doping level of the sample. As the doping concentration increases, more excitons are transformed into charged excitons which decay primarily through non-radiative channels, eventually leading to the decrease of PL intensity. For the readers' convenience, typical evolution curves of the PL intensity for bare $WS_2$ and encapsulated $WS_2$ are normalized and plotted together in Fig. 4(e). In consistence with the data shown in Fig. 2(d), we found that the encapsulated samples have a smaller time constant for the fast recovery process and a larger time constant for the slow recovery process than bare $WS_2$.

According to the experimental results presented above, we proposed a physical model which involves two mechanisms for the formation of fluorescence spatial hole burning effect. The first one is the EEA process. As shown schematically in Fig. 4(f), in the EEA process, one exciton is ionized by absorbing the energy from another exciton locating nearby. Due to the existence of electronegative sulfur vacancies which are well known to serve as centers of negative charges, holes generated from the EEA process are readily attracted and captured by these sulfur vacancies, leading to the increase of doping concentration in the sample.[37–40] As a result, more excitons are turned into charged excitons which decay primarily via non-radiative channels, causing the decrease of PL intensity. At the same time, irradiation by the high-power pumping laser beam causes desorption of electronegative $O_2$ and $H_2O$ molecules which originally adsorbed onto the surface of monolayer $WS_2$ from the surrounding atmospheric environment. This will also lead to the increase of doping concentration and decrease the PL intensity. As a result of the combined action of these two mechanisms, the fluorescence spatial hole burning effect emerges.

For the recovery process of the spatial hole burning effect, there are also two mechanisms involved. The first one is related to the release of those trapped holes. Indeed, the continuous irradiation of the weak probe beam provides the activation energy required for the dissociation of holes trapped by sulfur vacancies, leading to the release of the trapped holes. This causes de-doping of the sample and therefore increases the PL intensity. This mechanism is responsible for the fast recovery process we observed. The second mechanism is related to the re-adsorption of gas molecules. After the high-energy pumping beam is stopped, the electronegative $O_2$ and $H_2O$ molecules will re-adsorb onto the sample surface, which also leads to de-doping of the sample and increases the PL intensity. The second mechanism is responsible for the slow recovery process discussed in the previous text.

Using our physical model, the distinct time constants extracted from the bi-exponential recovery process for bare $WS_2$ and encapsulated $WS_2$ can be rationalized very well. As an ideal insulator with atomically flat surface (but without dangling bonds), the encapsulation using hBN could reduce the number of sulfur vacancies. Correspondingly, the number of holes trapped by sulfur vacancies decreases as well. Therefore, the release of trapped holes will be faster. This explains why hBN-encapsulated samples have a smaller time constant for the fast recovery process than bare $WS_2$. On the other hand, the encapsulation using hBN isolates $WS_2$ from the atmosphere. This prevents the re-adsorption of electronegative $O_2$ and $H_2O$ molecules. Therefore, hBN-encapsulated $WS_2$ has a larger time constant for the slow recovery process than bare $WS_2$.

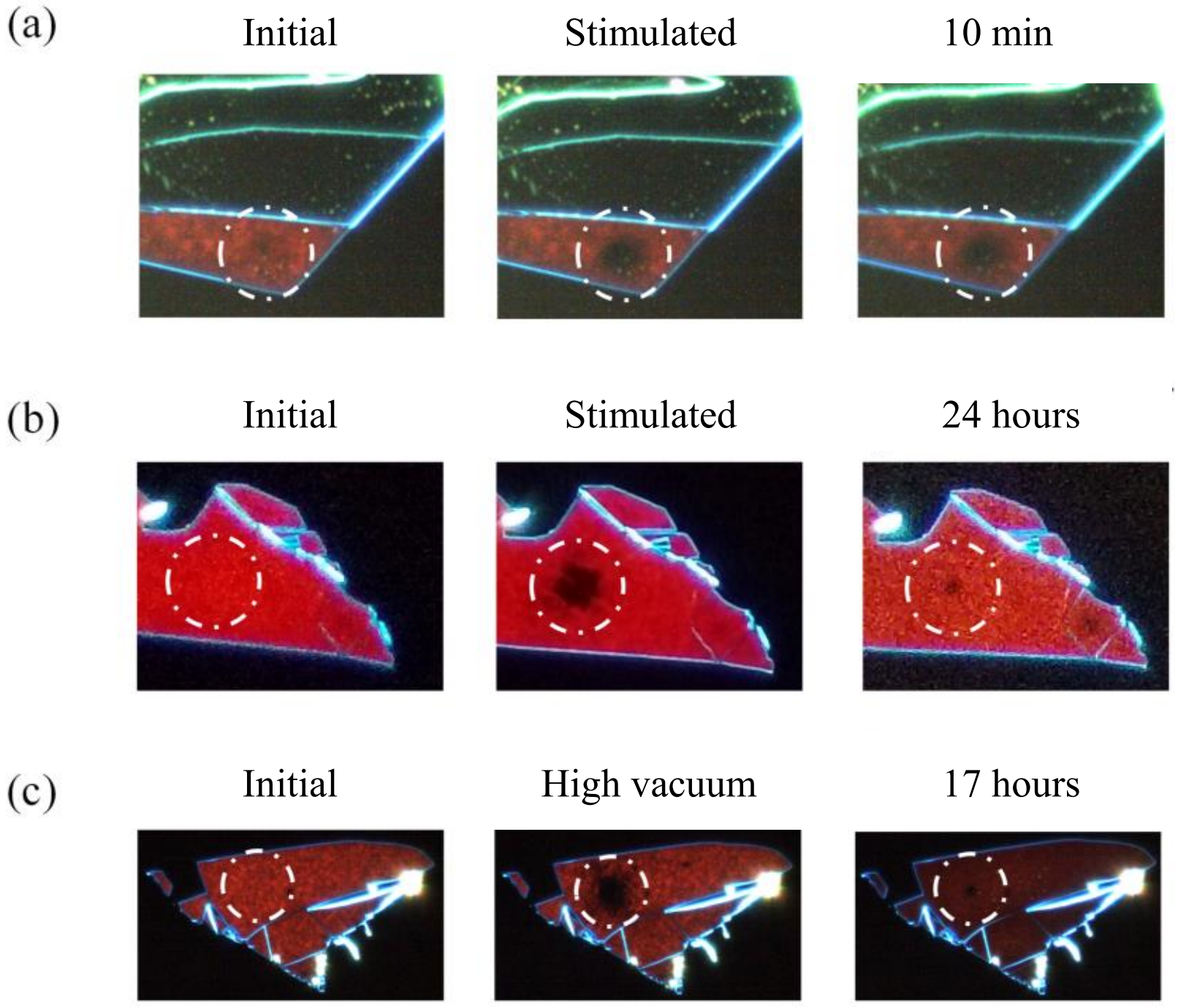

**Fig. 5. Dark field PL imaging of the fluorescence spatial hole burning effect and its recovery process.** (a) Dark field PL images of monolayer $WS_2$ when it is kept in a dark box under ambient conditions. Image denoted by "Initial": dark field PL image when the sample is excited by the weak probe beam only. Image denoted by "Stimulated": dark field image when the sample is excited by both probe and pumping beams simultaneously. Right image: dark field image when the sample has been exposed to the atmosphere for 10 minutes. (b) Dark field PL images of another monolayer $WS_2$ flake when it is kept in a dark box under ambient conditions. Right image: dark field image when the sample has been exposed to the atmosphere for 24 hours. (c) Dark field PL images of monolayer $WS_2$ when it is kept in a dark box under high vacuum condition ($1.4 \times 10^{-4}$ Pa). Right image: dark field image when the sample has been kept in high vacuum for 17 hours.

Having revealed the mechanism for fluorescence spatial hole burning effect and its bi-exponential recovery process, the next question is which mechanism makes the major contribution? To answer this question, we carried out dark field PL imaging experiments. Typical results are shown in Fig. 5. In all such measurements, the samples are kept in a dark box so that light-assisted release of trapped holes can be prevented. Fig. 5(a) shows the dark field PL images of monolayer $WS_2$ when it is exposed directly to the atmosphere. As one can see, with the application of the high-power pumping beam (the middle image denoted by "Stimulated"), a dark spot (*i.e.*, a "hole") appears immediately. This fluorescence hole remains almost unchanged when the sample has been exposed to the atmosphere for ~10 minutes. To further study the recovery process, we repeated our measurements using another $WS_2$ flake. As shown in Fig. 5(b), a fluorescence hole can again be observed clearly when the high-power pumping beam was applied. After the pumping beam was stopped, we kept the sample in a dark box and exposed it to the atmosphere for ~24 hours. Interestingly, the fluorescence hole disappeared completely. As light-assisted release of trapped holes has been prevented, there are two possible mechanisms for the recovery of the fluorescence hole. The first possibility is the re-adsorption of electronegative $O_2$ and $H_2O$ molecules which leads to de-doping of the sample. The second possibility is the natural release of trapped holes without the assistance of light irradiation. To clarify these, we continue dark field PL imaging experiments by placing the sample in a high vacuum chamber ($1.4 \times 10^{-4}$ Pa) and keeping it in a dark box. Typical results are shown in Fig. 5(c). As one can see from the middle image, clear fluorescence hole can again be observed. Surprisingly, by keeping it in high vacuum for ~17 hours, the fluorescence hole again disappeared. As re-adsorption of electronegative $O_2$ and $H_2O$ molecules is excluded, the only reasonable mechanism for the recovery is the spontaneous release of trapped holes. Based on these results, we could also know that the release of trapped holes makes the major contribution to the recovery of fluorescence spatial hole burning, while re-adsorption of electronegative molecules makes minor contribution. Here, it's also very exciting to note that the storage of holes by electronegative sulfur vacancy defects can persist for more than ten hours without light irradiation. Such ultralong-term storage of holes could find potential applications in optoelectronic devices.

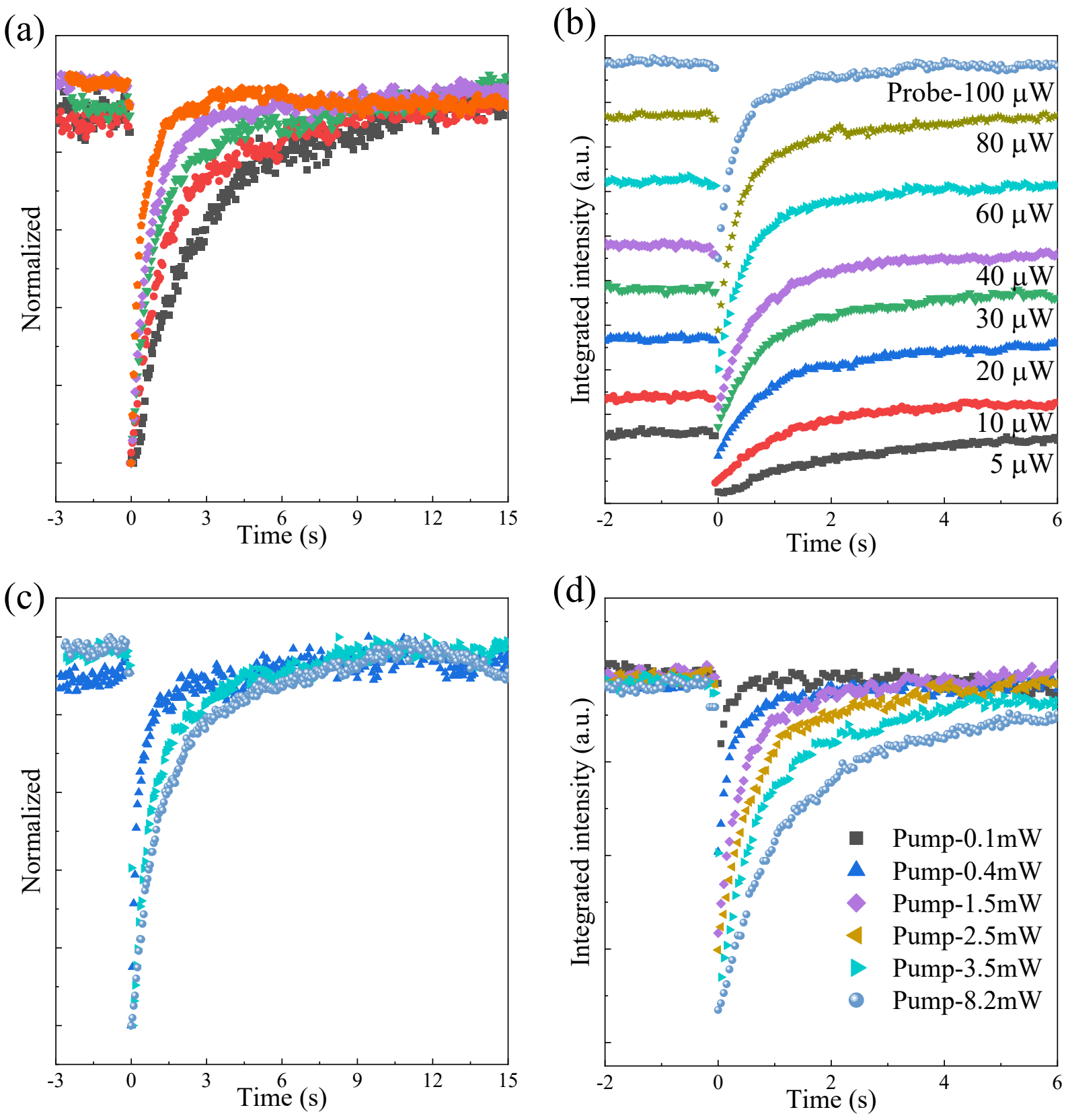

**Fig. 6. Power dependence of the fluorescence spatial hole burning effect and its recovery process.** (a) Normalized PL intensity as a function of time under different powers of the probe beam. Power of the pumping beam was fixed at 3 mW. (b) Intensity evolution curves without normalization, corresponding to those data shown in (a). (c) Normalized PL intensity as a function of time under different powers of the pumping beam. Power of the probe beam was fixed at 20 μW. (d) Intensity evolution curves without normalization, corresponding to those data shown in (c).

Next, we turn to the possible optical control of the fluorescence recovery process. Fig. 6 shows the power dependence of fluorescence spatial hole burning effect and its recovery process. Fig. 6(a) shows the normalized PL intensity as a function of time under our dual-beam pumping scheme. The corresponding data without normalization is plotted in Fig. 6(b). In this set of measurements, power of the pumping beam was fixed at 3 mW. Meanwhile, power of the probe beam was increased gradually from 5 μW to 100 μW. As one can see, with increasing power of the probe beam, the intensity dip at the zero moment becomes deeper. More importantly, it is obvious that the recovery rate of fluorescence becomes faster and faster. As the recovery is driven by the release of trapped holes, faster recovery rate means faster release rate of trapped holes. This tells us explicitly that the release of trapped holes in $WS_2$ can be fully controlled by laser irradiation. As a comparison, Figs. 6(c) and 6(d) show the intensity evolution curve where the pumping beam power was increased gradually while the probe beam was fixed at 20 μW. Again, the intensity dip at the zero moment becomes deeper with the pumping beam power, as higher power causes stronger EEA and therefore stronger photo-doping effect. In sharp contrast to those data shown in Figs. 6(a) and 6(b), the recovery rate becomes slower at higher pumping beam power. This is because more EEA-induced holes are trapped by sulfur vacancy defects at higher pumping beam power. Therefore, it requires more time for these trapped holes to be released. This consistency again gives strong support to our proposed model.

Based on the discussion above, we could also explain why the time constant for the fast recovery process and the FWHM of the fluorescence hole show saturation behaviors in their power dependent curves as shown in Figs. 3(c) and 3(d). According to our model, the fast recovery process corresponds to the release of trapped holes. Releasing more trapped holes requires more time. However, it should be noted that the number of intrinsic sulfur vacancies is fixed. When sulfur vacancy defects are all occupied by holes, time constant for the fast recovery process will saturate. Meanwhile, as all sulfur vacancy defects are occupied, photo-doping, which is due to the capturing of holes by sulfur vacancy defects, will stop. As a result, the fluorescence spatial hole burning, which mainly comes from EEA-induced photo-doping, reaches saturation.

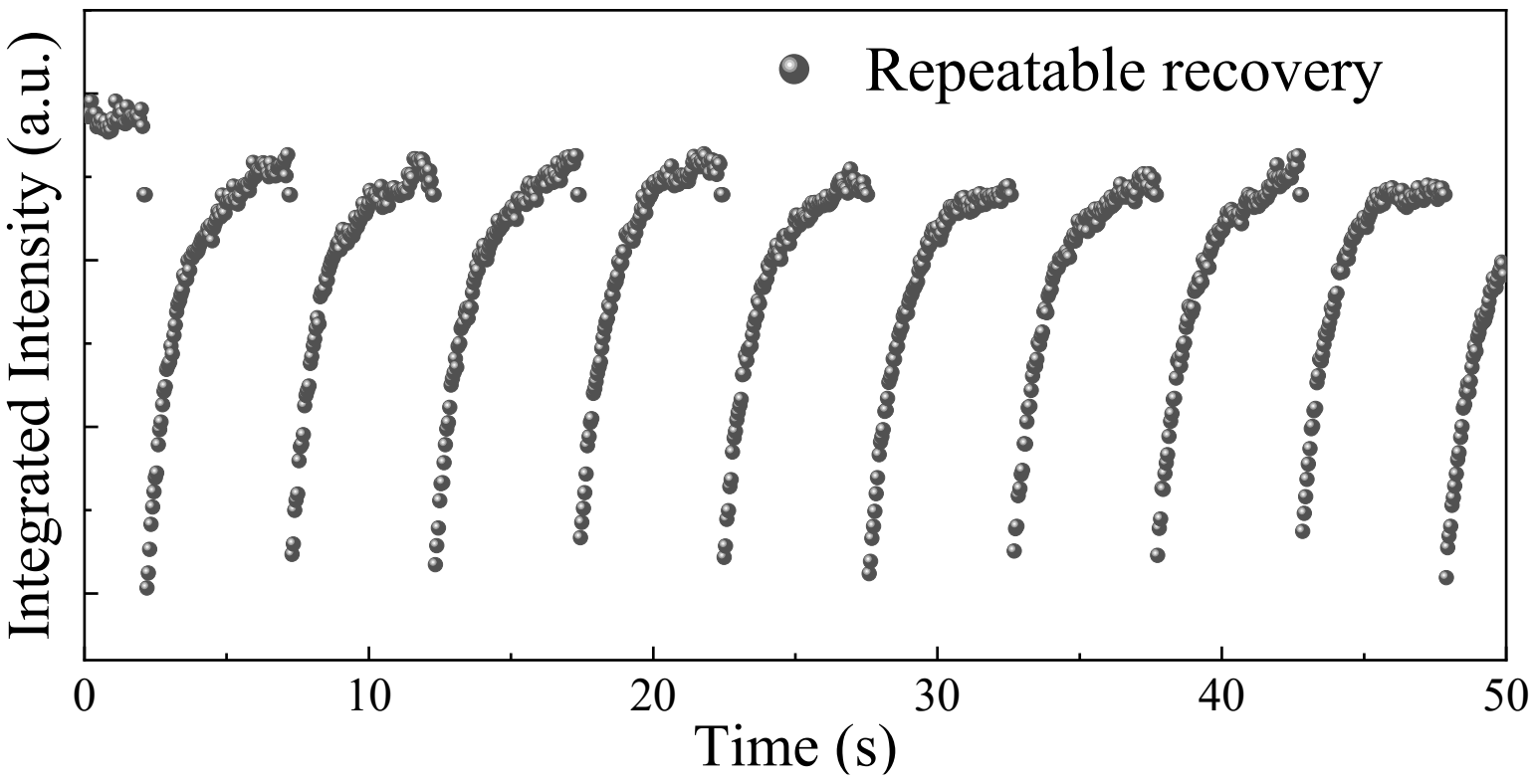

**Fig. 7. Repeatability of the fluorescence spatial hole burning effect and its recovery process.** The 405 nm pumping beam was applied to the sample periodically (Power: ~3 mW; single irradiation time: ~50 ms; period: ~5 s). The 532 nm probe beam (Power: ~20 μW) irradiated the sample continuously.

Finally, we would like to discuss the repeatability of the fluorescence spatial hole burning effect and its recovery process. Fig. 7 shows the integrated PL intensity as a function of time where the high-power pumping beam was applied periodically while the weak probe beam kept exciting the sample. As one can see, the fluorescence quenching and recovery process is highly repeatable. This highlights the broad potential application of TMDCs in the field of controllable photo-doping.

## Conclusions

In summary, we report systematic studies on the fluorescence spatial hole burning effect and its recovery process in monolayer $WS_2$. We revealed that the fluorescence spatial hole burning effect originates from photo-doping, which is induced by EEA process and the capturing of holes by sulfur vacancy defects. The recovery process of the fluorescence spatial hole burning effect was revealed to consist of two components: the fast recovery process comes from the release of trapped holes and the slow process is related to the re-adsorption of electronegative gas molecules. Based on our experimental data, we found that sulfur vacancy defects in monolayer TMDCs can achieve ultralong-term storage of holes (more than 10 hours). Moreover, the release of such trapped holes can be fully controlled by an external weak light beam. Our results not only contribute to a deeper understanding of the optical properties of monolayer TMDCs, but could also facilitate their device applications in the field of controllable photo-doping.

Acknowledgments. This work was supported by the National Natural Science Foundation of China (Grant No. 12274159), and the National Key Research and Development Program of China (Grant No. 2022YFA1602700).